\documentstyle[12pt]{article}

\makeatletter

\@addtoreset{equation}{section}
\def\section{\@startsection {section}{1}{\z@}{-3.5ex plus -1ex minus
 -.2ex}{2.3ex plus .2ex}{\large\bf}}
\def\subsection{\@startsection{subsection}{2}{\z@}{-3.0ex plus%
 -1ex minus -.2ex}{1.5ex plus .2ex}{\bf}}

\advance \voffset by -1.0in
\advance \hoffset by -0.6in
\def\cd{\!\cdot\!}

\def\bea{\begin{eqnarray}}
\def\eea{\end{eqnarray}}

\def\bphi{{\mbox{\boldmath $\phi$}}}
\def\bvphi{{\mbox{\boldmath $\varphi$}}}

\def\bp{{\mbox{\boldmath $p$}}}

\def\bq{{\mbox{\boldmath $q$}}}
\def\bn{{\mbox{\boldmath $n$}}}

\def\bx{{\mbox{\boldmath $x$}}}

\def\bR{{\mbox{\boldmath $R$}}}

\def\bS{{\mbox{\boldmath $S$}}}

\def\b0{{\mbox{\boldmath $0$}}}

\begin{document}
\baselineskip 18pt
\parskip 6pt
\begin{flushright}
DTP 94-25

hep-th/9410256
\end{flushright}

\begin{center}
{\LARGE Dynamics of Baby Skyrmions}

\vspace{1cm}
\baselineskip 24 pt
{\Large
B.M.A.G. Piette$^1$, B.J. Schroers$^2$ and  W.J. Zakrzewski$^3$\\
Department of Mathematical Sciences\\
South Road\\
Durham DH1 3LE\\
United Kingdom}
\\
\baselineskip 18pt
$^1${e-mail: {\tt b.m.a.g.piette@durham.ac.uk}}\\
$^2${e-mail: {\tt b.j.schroers@durham.ac.uk}} \\
$^3${e-mail: {\tt w.j.zakrzewski@durham.ac.uk}}
\vspace{1.2cm}

{\bf Abstract}

\end{center}

{\small

Baby Skyrmions are topological solitons in a (2+1)-dimensional  field
theory which resembles  the  Skyrme model in important respects.
We apply some of the  techniques and approximations commonly used in
discussions of the Skyrme model to  the dynamics of baby Skyrmions and directly
test them against numerical simulations.
 Specifically  we study the
effect of spin on the shape of a single  baby Skyrmion, the dependence of the
forces
between
 two baby Skyrmions on the baby Skyrmions' relative orientation and
the forces between two  baby Skyrmions when one of them is   spinning.   }

\section{Introduction}

The goal of this paper is to study  the dynamics of  solitons  in a
(2+1)-dimensional   version of the  Skyrme  model.
By a soliton we  mean a localised, finite-energy solution of a non-linear
field  theory. The Skyrme model is a non-linear field theory for pions
in 3+1 dimensions   with  soliton solutions called Skyrmions \cite{Skyrme}.
Suitably quantised Skyrmions are models for physical baryons.
Skyrme's theory is non-integrable and therefore progress in understanding
Skyrmion dynamics has depended on  numerical simulations,  approximation
schemes or a combination of both. This approach has been quite successful
in the study of  static soliton solutions in Skyrme's theory \cite{BT}
\cite{LM}. However, the interactive dynamics of two or more Skyrmions, which
one needs to understand in order to extract the Skyrme model's  predictions for
the nuclear
two-body problem, is more difficult to describe.
Certain  scattering
processes of two Skyrmions has been simulated  numerically \cite{VWWW} but the
variety of
possible initial conditions that one could consider is so  great that
it seems impossible to get an overall picture of the scattering behaviour
from just a few processes.
On the other hand,  various approximations have been used which typically
involve truncating the field theory with infinitely many degrees of freedom
to a finite-dimensional dynamical system \cite{M2}. Some approximations have
become
widely accepted without, however, having been tested against  numerical
simulations of the full theory.  Here  we will apply many of the
 concepts and   approximations developed in the Skyrme model to our model
and directly compare  them  with  numerical simulations.

Our solitons  are
 exponentially localised in space, a property shared by Skyrmions
when  the physical pion mass is included in the Skyrme  model.
In our model  a soliton has a fixed size  but arbitrary
position and orientation. In two spatial  dimensions this  corresponds to
three degrees of freedom, two for the soliton's position and
one angle to describe its orientation. This should again be compared
with Skyrmions, which
 have
a definite size  and six degrees of freedom, three  giving its position
in space and  three  parametrising  its  orientation.
The long-range interaction  behaviour of our solitons resembles
that of Skyrmions in important respects: in both cases the asymptotic forces
between two solitons depend on their separation and their
relative orientation and  are of the dipole-dipole type. Furthermore, there
is a bound state of two solitons with a toroidal energy distribution in
both cases. When the solitons
are orientated so that the forces are most attractive and then released from
rest they scatter through  the toroidal configuration and emerge at
$90^{\circ}$ degrees relative to initial direction of their motion.

Because of all these similarities we call our solitons  baby Skyrmions.
This term has been used  quite widely to describe solitons in 2+1 dimensions
which resemble Skyrmions in certain respects. However, in all
 the models studied so far the moment of inertia  for the rotation of a
single soliton  is infinite so that the rotational degrees of freedom
are not dynamically relevant. Yet the rotational degrees of freedom
are crucial in Skyrmion dynamics.  In this paper we are therefore particularly
interested
in those aspects
of our model  which  depend on the
 solitons' orientation.

After a  description of our model and a quick review of its static solutions,
discussed in detail in \cite{PSZ}, we
 focus on the following questions.
Are there exact solutions of the field equations representing spinning
baby Skyrmions? How does  a baby Skyrmion change its shape when it spins?
How do the asymptotic forces between two baby Skyrmions depend on their
relative orientation? What are the forces between two baby Skyrmions when
one of them is spinning?

While the questions we  investigate here are largely motivated
by the (3+1)-dimensional Skyrme model we  should emphasise that our
model is also of interest in
the large and growing area of (2+1)-dimensional soliton
phenomenology. The solitons in our model have a number of properties which
are novel in this context.

\section{The Model}
The basic field of our model is a map
\bea
\bphi : \mbox{M}^3 \mapsto  S_{\mbox{\tiny iso}}^{2}
\eea
where M$^3$ is three-dimensional Minkowski space with the metric
diag$(1,-1,-1)$. We set the speed of light to $1$ and write    elements of
M$^3$ as $(t,\bx)$, where $\bx$ is a 2-vector with coordinates
$x^i$, i=1,2,  also sometimes denoted by $x$ and $y$,  We use the
notation $x^{\alpha}$, $\alpha=0,1,2$, to refer to both the time and spatial
components of $(t,\bx)$, so  $t=x^0$.
The target space $S^2{\mbox{\tiny iso}}$ is the 2-sphere of unit radius
embedded in
euclidean 3-space with  the Riemannian metric induced by that embedding.
Here the suffix `iso' is used  to emphasise the analogy with the target
space in the Skyrme model, which is often referred to as iso-space. The field
$\bphi$ is a scalar field with  three components $\phi_a$, $a = 1,2,3$,
satisfying the constraint  $\bphi\cd \bphi = \phi_1^2 + \phi_ 2^2 +
\phi_3^2= 1$ for all
$x\in $ M$^3$.
Our  Lagrangian density, the static part of which was first considered   in
\cite{Tigran} and discussed further in \cite{PSZ},
 is
\bea
\label{Lag}
{\cal L} =  F\left( {1 \over 2}
\partial_{\alpha} \bphi \cd \partial^{\alpha} \bphi
- {\kappa^2 \over 4}
(\partial_{\alpha} \bphi \times \partial_{\beta} \bphi) \cd
(\partial^{\alpha} \bphi \times \partial^{\beta} \bphi)
- \mu^2 (1-\bn\cd\bphi) \right),
\eea
where $\bn =(0,0,1)$ and $\partial_{\alpha}={\partial /\partial x^{\alpha}}$.
The constants $F$, $\kappa$ and $\mu$ are free parameters: $F$ has the
dimension energy and $\kappa$ and $\mu$ have the dimension length.
 It is useful to  think of  $F$ and $\kappa$ as natural units of  energy and
length respectively and of $\mu$ as a second length scale in our model. Here we
fix
our units of energy and length  by setting $F=\kappa=1$.
We have already  set the speed of light to $1$, so  we   use `geometric' units
in which all
physical quantities are dimensionless. Thus we cannot set $\mu$ to $1$
by a choice of units, and we will fix its value later after we have
discussed its significance. Note also that  Planck's constant
$\hbar$ will be some number, but not necessarily equal to $1$.
The first term  in (\ref{Lag}) is familiar from  $\sigma$-models whose soliton
solutions
have been studied extensively \cite{sigma}. The second term, fourth order
in derivatives, is the
analogue of the Skyrme term in the  usual Skyrme model.
In \cite{PSZ} we explained in more precise terms  in which sense this analogy
holds
by  appealing to a general geometric framework
for the Skyrme model due to Manton \cite{M}.
 Finally, the last term does not contain any derivatives and is often
referred to simply as a potential. In  three
(spatial) dimensions the Skyrme term is necessary for the existence
of soliton solutions but  the inclusion of a potential is optional
from the mathematical point of view. Physically, however, a potential
of a certain form is required to give the pions a mass \cite{AN}.
By contrast, in two dimensions  a potential must be included
in the  above Lagrangian
in order to obtain soliton solutions.

To see this,
and to understand the Lagrangian  $L= \int {\cal L} d^2 x $ better,
we write it in the usual form
$L=T-V$. Here $T$ is the kinetic energy
\bea
T = \int  {1\over 2} \dot \bphi \cd \dot \bphi
+{1\over 2}(\dot\bphi \times \partial_i\bphi)\cd (\dot\bphi \times \partial_i
\bphi) d^2 x,
\eea
where the dot denotes differentiation with respect to time, and
$V$ is the potential energy
\bea
\label{V}
V = \int
 {1\over 2} \partial_i\bphi \cd \partial_i \bphi
+{1\over 4}(\partial_i\bphi \times \partial_j\bphi)\cd (\partial_j \bphi \times
\partial_i
\bphi) +\mu^2 (1-\bn\cd\bphi) d^2 x .
\eea
We are only interested in fields with finite potential energy and
we therefore impose the boundary condition
\bea
\label{bound}
\lim_{|\bx| \rightarrow \infty} \bphi(t,\bx) = \bn
\eea
for all $t$.
As  a result we can formally compactify physical space to a 2-sphere
$S^2_{\mbox{\tiny space}}$ and regard the fields $\bphi$ at a fixed
time  $t$ as maps from $S^2_{\mbox{\tiny space}}$ to $S^2_{\mbox{\tiny iso}}$
with an associated  integer degree.
The analytical formula for the degree is
\bea
\label{degree}
\mbox{deg}[\phi] = {1\over 4 \pi}\int \bphi \cd \partial_1 \bphi \times
\partial_2 \bphi d^2x .
\eea
The degree is a homotopy invariant of the field $\bphi$ and therefore
conserved during time evolution. Moreover, it gives a useful lower bound
on the potential energy, the Bogomol'nyi bound
\bea
\label{Bog}
V[\bphi] \geq 4\pi|\mbox{deg}(\bphi)|.
\eea

It is well known  that  the usual $\sigma$-model Lagrangian, where both the
Skyrme term
and the potential are omitted,
is scale invariant. In particular, this means that a soliton
solution of the $\sigma$-model can have arbitrary size.
The inclusion of the Skyrme term breaks the scale invariance and it follows
from a simple scaling argument that the potential energy contributed by the
Skyrme term   can be lowered by increasing the scale of a  configuration.
Thus, in order to obtain stable solutions it is necessary to include a
potential
which, on its own, would favour small scales. It follows that soliton solutions
in a Lagrangian with both Skyrme term and potential have a  definite  size.

 The precise form of the potential does not matter in such
scaling arguments   and
potentials other than the one considered here have been studied in
the literature \cite{pot}.
Our potential is the analogue of the potential usually chosen in the
Skyrme model.
 It contains  a constant $\mu$  which, in the language of quantum field theory,
can be interpreted as the inverse Compton wavelength
of the mesons in our model.  To see this it is best to turn to
the equations of motion.

The Euler-Lagrange equations for  the Lagrangian $L$
are
\bea
\label{EL}
\partial^{\alpha}\left(\bphi \times \partial_{\alpha}\bphi
-\partial_{\beta}\bphi (\partial^{\beta}\bphi\cd \bphi \times \partial_{\alpha}
\bphi) \right)
= \mu^2 \bphi \times \bn.
\eea
One simple solution is given by  $\bphi(t,\bx) = \bn $. It has degree  zero
and is called the vacuum configuration.
For a physical interpretation of our Lagrangian it is useful to study
the  equation obeyed by small fluctuation around the vacuum
configuration. Decomposing $\bphi$ into a component parallel to the vacuum
and a component $\bvphi$ orthogonal to it
\bea
\label{asyf}
\bphi = \sqrt{1-\bvphi^2}\bn + \bvphi \approx
\bn + \bvphi + {\cal O}(\bvphi^2), \qquad \bvphi\cd\bn =0,
\eea
one checks that the linearised equation for $\bvphi$ is the massive Klein
Gordon
equation
\bea
(\Box + \mu^2 )\bvphi = 0
\eea
where $\Box =\partial_{\mu}\partial^{\mu}$ is the wave operator in 2+1
dimensions.
In the language of perturbative quantum field theory, where one quantises
small fluctuations around the vacuum, the scalar fields $\varphi_1$ and
$\varphi_2$
therefore correspond to  scalar particles or mesons of mass $\hbar \mu$.
 We now understand $\mu$ well enough to fix its value. Our choice is again
dictated by the
desire to reproduce  important features of the Skyrme model. There, as
in real nuclear physics, the Compton wavelength of the pion is  of the
same order as (in fact,  slightly larger than) the size of a Skyrmion.
Thus, we want to tune $\mu$  in our model so that the energy distribution of
its basic soliton solution  is concentrated  in a region of diameter $\approx
1/\mu$. By trial and error we find that this is  the case when we set
$\mu^2=0.1$, which we do for the rest of this paper.

For later use we note a conservation law  that can be read off
immediately from the equation of motion (\ref{EL}). Taking the
scalar product with $\bn$ on both sides of the equation we find
that the current
\bea
\bn\cd \bphi \times \partial_{\alpha}\bphi -(\bn\cd \partial_{\beta}\bphi)
(\partial^{\beta}\bphi\cd \bphi \times \partial_{\alpha} \bphi)
\eea
has vanishing divergence. The symmetry that leads to this conservation
law is $SO(2)$  rotations of the field $\bphi$ about $\bn$, which can be
written in terms of an angle $\chi$ as
\bea
(\phi_1,\phi_2,\phi_3)\mapsto (\cos \chi\ \phi_1 +\sin\chi\phi_2,
-\sin\chi\phi_1 + \cos\chi \phi_2,\phi_3).
\eea
 We call such a transformation an iso-rotation;  the corresponding conserved
quantity  is called isospin and denoted by $I$:
\bea
I = \int
\bn\cd \dot\bphi \times \bphi +(\bn\cd\partial_i\bphi) (\partial_i\bphi\cd
\dot\bphi \times \bphi) d^2 x.
\eea
For  a  systematic discussion of the symmetries of our model
we refer the reader  to  \cite{PSZ}, but here we need only  note that both
the Lagrangian $L$ and the degree (\ref{degree}) are invariant under
simultaneous reflections in space and iso-space
\bea
\label{Px}
P_x: (x,y)\mapsto (-x,y)\quad\mbox{and}\quad (\phi_1,\phi_2,\phi_3)
\mapsto (-\phi_1,\phi_2,\phi_3).
\eea
Later we will also make use of the invariance of  both the degree and
the Lagrangian under the combination of $P_x$ with a
 with a rotation by $\pi$ in both space and iso-space:
\bea
\label{Py}
P_y: (x,y)\mapsto (x,-y)\quad\mbox{and}\quad (\phi_1,\phi_2,\phi_3)
\mapsto (\phi_1,-\phi_2,\phi_3).
\eea

  \section{Static Solutions Revisited }

Time-independent solutions of the equations  of motion (\ref{EL}), which
are stationary points of the potential energy  functional $V$ (\ref{V}), were
studied in detail in \cite{PSZ}. We briefly recall those results which are
relevant here.
 An important class of static solutions
of the equations of motion consists of fields which are invariant under   the
group of simultaneous  spatial rotations by some angle $\alpha \in [0,2\pi)$
and  iso-rotations by $-n\alpha$, where $n$ is a non-zero integer. Such fields
are of the form
\bea
\label{hedge}
\bphi(\bx) = \left(
\begin{array}{l}
\sin f(r) \cos (n\theta-\chi)\\
 \sin f(r) \sin(n\theta - \chi)\\
 \cos f (r)
\end{array} \right),
\eea
where $(r,\theta)$ are polar coordinates in the $\bx$-plane
 and $f$ is  function satisfying certain boundary conditions to be specified
below. The angle $\chi$ is also arbitrary,
but fields with different $\chi$ are related by an iso-rotation and therefore
degenerate in energy. Thus we  concentrate on the standard fields where
$\chi =0$.
Such fields are the analogue of the `hedgehog'  fields
in the Skyrme model and were also studied in \cite{Tigran} for different values
of  $\mu^2$.

The  function $f$, which is called the profile function, has to satisfy
\bea
\label{origin}
f(0)= m\pi, \quad m\in \mbox{\bf Z},
\eea
 for the field (\ref{hedge}) to be regular at
the origin; to satisfy  the boundary condition (\ref{bound})  we set
\bea
\label{infinity}
\lim_{r\rightarrow \infty} f(r) = 0.
\eea
Here we will only be interested in  profile functions where $m=1$
(the situation for general $m$ is discussed in \cite{PSZ}).
One then finds  that the degree of the field (\ref{hedge}) is
\bea
\mbox{deg}[\bphi] =n.
\eea
For a field of the form   (\ref{hedge}) to be  a stationary point of the energy
functional $V$, $f$ has to satisfy the Euler-Lagrange equation
\bea
\label{rad}
\left(r + {n^2 \sin^2 f\over r}\right)f'' +\left(1-{n^2 \sin^2f\over r^2} +
{n^2 f'\sin f \cos f \over r}\right)f' -{n^2\sin f \cos f \over r} - r\mu^2\sin
f
=0 \nonumber \\
{}.
\eea
It was shown in \cite{PSZ} that the hedgehog  fields (\ref{hedge}) with
$n=1$ and $n=2$ and  profile functions satisfying the equation above for
those values of $n$ are the absolute minima of $V$ amongst all field
of degree $1$ and $2$ respectively. We write $\bphi^{(1)}$ and $\bphi^{(2)}$
for those fields in standard iso-orientation, i.e. with $\chi=0$ in
(\ref{hedge}), and denote their profile functions by $f^{(1)}$ and $f^{(2)}$.
Translations in physical space and iso-rotations
act non-trivially on $\bphi^{(1)}$ and $\bphi^{(2)}$, so there is
 a three-dimensional family of minima of the energy functional for both $n=1$
and $n=2$.
We call any field obtained by translating and iso-rotating the
field  $\bphi^{(1)}$ a baby Skyrmion. Baby Skyrmions are the basic
solitons of our model; as promised in the introduction  they have  three
degrees  of freedom: two translational
and one rotational. For the total energy, or mass, of a baby Skyrmion we find
$1.564 \cd 4\pi$; the energy density is rotationally  symmetric and peaked at
the baby Skyrmion's centre (where $\bphi^{(1)}=(0,0,-1)$). The field
$\bphi^{(2)}$ (and all those obtained
by translating and rotating it)  may be thought of as a bound state
of two baby Skyrmions and is described in detail in \cite{PSZ}. The energy
density is again rotationally symmetric
but peaked at a distance $r\approx 1.8$ from the centre. The  mass is $2.936
\cd 4\pi$.

In subsequent sections we will be interested in the field of a spinning baby
Skyrmion and the  forces between two baby Skyrmions. We  recall some
simple observation concerning the  asymptotic behaviour of $\bphi^{(1)}$ from
\cite{PSZ} which
are the basis of  a remarkably accurate model for  the  dynamical phenomena  we
will then
encounter.
For large $r$, and hence small $f$,  the equation (\ref{rad}) for $n=1$
simplifies to the modified Bessel equation
\bea
f'' + {1\over r} f' - ({1\over r^2} + \mu^2)f=0.
\eea
A solution of this equation which tends to zero at $r=\infty$ is the modified
Bessel function $K_1(\mu r)$. Thus, the profile function  $f^{(1)}$ of
(\ref{rad}) is proportional to $K_1$ for large $r$ and we can write
\bea
\label{asy}
f^{(1)}(r) \sim {p \mu \over2 \pi}K_1(\mu r),
\eea
where $p$ is a constant which we will
interpret further below. Since the modified Bessel function   has the
asymptotic behaviour
\bea
\label{asympt}
K_1(\mu r) \sim \sqrt{\pi \over 2 \mu r} e^{-\mu r}\left( 1+ {\cal O}({1\over
\mu r})\right)
\eea
 the leading term in an asymptotic expansion of  $f^{(1)}$ is proportional to
$e^{-\mu r}/ \sqrt r$,
which shows in particular  that the (potential) energy distribution of the
field $\bphi^{(1)}$ is exponentially
localised.
Most of the analytical results in this paper are based on the observation,
made
in \cite{PSZ}, that  the asymptotic field $\bvphi^{(1)}$ of $\bphi^{(1)}$
(defined as in (\ref{asyf})) can be  interpreted in terms of dipole fields.
This can be seen as follows. For large $r$
 we  approximate
$\sin f^{(1)} \sim f^{(1)}$ and $\cos f^{(1)} \sim 1$ and, using the asymptotic
expression (\ref{asy}) we write the
 field $\bvphi^{(1)}$ as
\bea
\bvphi^{(1)}(\bx) = {p\mu\over 2\pi} K_1(\mu r) \left(
\begin{array}{l}
\cos (\theta-\chi) \\
 \sin(\theta -\chi) \\
0 \end{array} \right).
\eea
Or, introducing the orthogonal  vectors
\bea
\bp_1 = p(\cos\chi,\sin\chi) \qquad \bp_2= p(-\sin\chi, \cos \chi)
\eea
and $\hat \bx =\bx/r$, we can write
\bea
\varphi^{(1)}_a(\bx)={\mu \over  2\pi} \,\bp_a\cd\hat\bx K_1(\mu r)
= -{1\over 2 \pi}
\bp_a\cd\nabla K_0(\mu r)
 \qquad a=1,2.
\eea
However, since  the Green function  of the static
Klein-Gordon equation is
$K_0(\mu r)$,
\bea
\label{Green}
(\Delta -\mu^2)K_0(\mu r)= -2\pi \delta^{(2)}(\bx),
\eea
we have
\bea
(\Delta -\mu^2)\varphi_a^{(1)}(\bx) = \bp_a\cd\nabla\delta^{(2)}(\bx)
\qquad a=1,2.
\eea
This equation leads to the interpretation of  the asymptotic field
$\bvphi^{(1)}$ as the field
produced by a doublet
of orthogonal dipoles, one for each of the components $\varphi^{(1)}_1$ and
$\varphi^{(1)}_2$, in a linear field theory, namely Klein-Gordon theory.
The strength of the dipole can be calculated from the asymptotic form
of $f^{(1)}$. One finds, by numerically solving the equation (\ref{rad}),
\bea
p = 24.16. \qquad
\eea
 Once this single number is calculated from the non-linear equation (\ref{rad})
much can be deduced  about the dynamics of baby Skyrmions  using only
the linearised equations of motion.

\section{Spinning Baby Skyrmions}

How does a soliton in two or three dimensions  change its shape when
it spins?  What is the interactive dynamics of several solitons when some of
them are spinning? From the point of view of particle physics these are very
natural
questions to ask. Yet there are surprisingly few soliton models in which
these questions has been addressed seriously and even fewer in which
satisfactory answers have been found. This is partly because the questions
do not make sense in some of the most popular  models. In  the much
studied abelian Higgs model \cite{vortices}, for example,  the soliton
solutions, called vortices, are fully characterised by their position and
have no rotational degrees of freedom. Lumps in the ${\bf CP}^1$ model,
on the other hand, can have an arbitrary orientation, but the moment of inertia
associated with changes in the orientation is infinite so
that the rotational degree of freedom  is dynamically frozen out.
There is a modification of the
${\bf CP}^1$ model \cite{Qlumps} in which the solitons, called Q-lumps,
necessarily spin, but  single soliton solutions  have infinite energy and
in configurations of several solitons all solitons have to rotate with
the same angular frequency. Thus, the effect of relative rotation cannot
be investigated.

In the Skyrme model, spin $1/2$ quantum states of a single Skyrmion
are models for physical nucleons, so the question of spinning Skyrmions
has naturally attracted a lot of attention. In the first paper
on this subject \cite{ANW}, it was assumed that  a Skyrmion would
rotate without changing its shape, and, to obtain the quantum states
corresponding to the proton, neutron and the $\Delta$-resonance,   it was
quantised as a rigid body. Although it was quickly pointed out that the
classical frequency at which a Skyrmion would have to rotate to have
spin $1/2$ is so large that centrifugal and relativistic effects
are important, there seems to be no quantitative analysis of these effects
in the literature. One reason for this is that, in three dimensions, a
rotating Skyrmion only has  axial symmetry (about the axis of rotation)
and is  not  of the $SO(3)$ symmetric hedgehog form of the static solution.
Thus, to find the
exact form of a spinning Skyrmion one needs to solve coupled non-linear
partial differential equations similar to the ones studied in \cite{BC}.

In our two-dimensional model, by contrast, there are  solutions describing
spinning
baby Skyrmions which are of the hedgehog form (\ref{hedge}). Thus we
can study the effect of rotation on the soliton's shape, its mass and its
moment
of inertia by simply solving ordinary differential equations.
It follows from the `principle of symmetric criticality' \cite{symcrit}
that we can find  time-dependent solutions  of the field
equations (\ref{EL}) by making the time-dependent hedgehog ansatz
\bea
\label{spin}
\bphi^{\omega}(t,\bx) =\left(
\begin{array}{l}
\sin f(r) \cos (\theta-\omega t)\\
 \sin f(r) \sin(\theta - \omega t)\\
 \cos f (r)
\end{array}
\right),
\eea
where $\omega$, an arbitrary real number, can be interpreted as the field's
angular frequency
and $f$ satisfies the boundary conditions (\ref{infinity}) and (\ref{origin})
with $m=1$. Then the field (\ref{spin})
is a solution of the Euler-Lagrange equation (\ref{EL})  if $f$ satisfies
the Euler-Lagrange equation obtained   from the restriction of the Lagrangian
 $L$ to fields of the form (\ref{spin}).
Explicitly this is
the ordinary differential equation
\bea
\label{radspin}
 (r + ({1\over r} - \omega^2 r)\sin^2 f)f''& +&(1-(\omega^2 +{1\over r^2})\
\sin^2f +
({1\over r} -\omega^2 r) f'\sin f \cos f )f'\nonumber \\
& -&({1\over r} -\omega^2 r)\sin f \cos f  - r\mu^2\sin f
=0.
\eea
We call  solutions of the form (\ref{spin}) spinning baby Skyrmions. The total
energy
of a spinning baby Skyrmion depends on $\omega$ and is given by
\bea
\label{mass}
M(\omega) = \pi \int  r \left( f'^2 +(\omega^2 + {1\over r^2})(1 + f'^2 )
 \sin^2 f + 2\mu^2(1-\cos f) \right)dr.
\eea
 $M(0)$ is just the mass  of a baby Skyrmion calculated earlier and shall
henceforth be denoted $M_0$.
To study the dependence of  $M$ and the field $\bphi^{\omega}$  on  $\omega$ we
need to distinguish two regimes, $\omega<\mu$
and $\omega > \mu$. This can be seen from the behaviour of the $f$ for large
$r$. In this limit $f$ is small and the equation
(\ref{radspin}) simplifies to
\bea
\label{spinBe}
f'' + {1\over r} f' - ({ 1\over r^2} + (\mu^2-\omega^2))f=0.
\eea
Thus, for $\omega <\mu$, $f$   decays exponentially
for large $r$ while for $\omega> \mu$ it is oscillatory.

It is interesting to  compare
the asymptotic field of a spinning baby Skyrmion with the field
produced by a doublet of scalar dipoles.  We then need to consider
time-dependent  dipoles and note here that the equation for  scalar fields
$\varphi_a$
produced by  a  rotating pair of orthogonal   dipoles
$\bp_a(t)$, $ a =1,2$,  is  the Klein Gordon equation with a time-dependent
dipole
source term:
\bea
\label{spindip}
(\Box + \mu^2)\varphi_a(t,\bx) = -\bp_a(t)\cd\nabla\delta^{(2)}(\bx).
\eea
If the dipoles rotate uniformly at constant angular velocity $\omega$
then $\ddot \bp_a = -\omega^2 \bp_a$ and the above equation  can be  solved
explicitly in terms of Bessel functions. We will do this
below for the two regimes $\omega <\mu $ and $\omega > \mu$ and compare the
results with the  asymptotic form of \nolinebreak $\bphi^{\omega}$.

\subsection{The case $\omega < \mu$}

If $\omega <\mu$, the equation  (\ref{spinBe}) is again  the modified Bessel
equation of first order.
As before we are interested in a solution $f$ which is exponentially small for
large $r$. Such a solution is
 asymptotically proportional to the  modified Bessel function
$K_1(\kappa r)$, where  $\kappa  = \sqrt{\mu^2 -\omega^2}$, i.e.
\bea
f \sim  {\kappa p \over 2\pi } K_1(\kappa r)
\eea
for some constant $p$.
Then the asymptotic field  $\bvphi^{\omega}$ of the rotating  baby Skyrmion
(\ref{spin}) with that   profile
function is
\bea
\bvphi^{\omega}(t,\bx) = {p\kappa\over 2\pi} K_1(\kappa r)\left(
\begin{array}{l}
\cos (\theta-\omega t ) \\
 \sin(\theta -\omega t) \\
0 \end{array} \right).
\eea
Thus, defining the time-dependent dipole moments
\bea
\label{dips}
\bp_1= p(\cos \omega t,\sin\omega t) \quad \mbox{and} \quad
\bp_2=p(-\sin\omega t,\cos \omega t),
\eea
  the first two components of   $\bvphi^{\omega}$ can be written
\bea
\varphi^{\omega}_a(t,\bx) = -{1\over 2 \pi} \bp_a(t)\cd \nabla K_0(\kappa
r)\qquad
 a=1,2.
\eea
One checks easily that these fields satisfy the linear equation
(\ref{spindip}).
 Thus, just as in the static case, the  asymptotic form  of the rotating
hedgehog solution (\ref{spin})
may be thought of as being produced by a rotating pair of orthogonal
dipoles.

We have solved the radial equation (\ref{radspin})
for various values of $\omega < \mu$. In figure  1 we plot the corresponding
profile functions. As $\omega$ approaches $\mu$ from below, the soliton's
energy distribution becomes more and more spread out,  which one may
interpret as  a centrifugal effect. Note, however,   that the initial gradient
of the profile functions varies very little and that most of
the change occurs in the tail, which is well described by the modified Bessel
function. This is the first  indication that one can understand many of the
features
 of a spinning baby Skyrmion in terms  of the field in the asymptotic
region where it (approximately)  obeys linear  equations.

Next we  want  to understand the dependence of a spinning baby
Skyrmion's mass on its frequency.  Recalling the
asymptotic formula (\ref{asympt}) we see that the energy distribution of  a
baby Skyrmion  spinning at $\omega < \mu$ is exponentially localised and that
the total mass $M(\omega)$ is finite.  At the critical angular velocity
$\omega  =\mu$, however,
the equation (\ref{spinBe}) is solved by $f=1/r$; the corresponding
baby Skyrmion is thus only power-law localised and  its mass is infinite.
For later use we note that the divergent part of the integral in the formula
(\ref{mass}) is
\bea
\label{divergent}
\pi \int  r \left( \omega^2
\sin^2 f + 2\mu^2(1-\cos f) \right)dr.
\eea
Due to the spreading of the energy density as $\omega $ approaches $\mu$
one needs to integrate  to ever larger values of $r$
when computing
 $M(\omega)$  numerically  using  the formula (\ref{mass}).  It is then more
practical to
find  constants $r_0$ and  $C$ such that for $r>r_0$  the profile function is
well approximated
by $C \exp(-\kappa r)/\sqrt r$. Then one
 integrates the energy density numerically from $0$ to $r_0$ and
performs the remaining integral analytically, using   $\sin^2 f \approx f^2$and
$\cos f\approx  1-f^2  / 2$.
We  plot the function $M(\omega)$ in figure 2.a).
It grows rapidly as $\omega \rightarrow \mu$ which is consistent with
our earlier observation that  $M(\mu)$ is infinite.

A further quantity of interest is the conserved charge $I$ discussed
at the end of section 2. For fields of the hedgehog form, where spatial
rotations and
iso-rotations are equivalent, we can interpret $I$ as the angular
momentum or spin of the field $\bphi$ and we denote it by $J$ in this
context.
One  finds that
\bea
\label{J}
J(\omega) = \omega \cd 2\pi \int r \sin^2 f ( 1 + f'^2) dr.
\eea
The quantity
\bea
\Lambda(\omega) = J(\omega)/\omega
\eea
may be interpreted as a moment  of inertia. Since
a spinning
baby Skyrmion changes its shape as  $\omega$ varies, the
corresponding moment of inertia changes, too.
As mentioned above this effect is customarily ignored in the
 Skyrme model.  To check the validity of this approximation
in our model  we define $\Lambda_0 = \Lambda(0)$, for later use.
Numerically we find $\Lambda_0= 2\pi\cd 7.558$

We have calculated $J(\omega)$ for various values of $\omega <\mu$
using the same technique as described above for the computation of $M(\omega)$.
Like $M$, $J$ diverges as $\omega \rightarrow \mu$ from below,
but one checks that  only the  term
\bea
\omega \cd 2\pi \int r \sin^2 f dr
\eea
is a  divergent integral   when $\omega =\mu$.
The divergence is of the same order as  that in (\ref{divergent})
and, comparing coefficients,  one is  lead to the asymptotic formula
\bea
\label{miracle}
M \sim  N + \mu J
\eea
for some constant $N$. In figure 2.b)  we plot the precise relation
between $M(\omega)$ and $J(\omega)$. Clearly the graph is well described
by the linear formula (\ref{miracle}) already for quite small values of
$\omega$. Note that a linear formula of this form holds  {\it exactly} for
Q-lumps
\cite{Qlumps}, where the constant $N$ is the Q-lumps topological
charge. One may interpret it in words as

``mass of a spinning soliton = constant + meson mass $\times$ angular
momentum''.

\noindent For small $\omega$ the  mass $M(\omega)$ depends quadratically
on $J(\omega)$ as one would expect for the rotation of a rigid body.
It is instructive to compare our exact results with the non-relativistic
rigid body formula
\bea
\label{rigid}
\tilde M = M_0 + {J^2\over 2\Lambda_0}.
\eea
We plot the graph of  $\tilde M(J)$   is figure  2.b)  as well.
Note that the rigid body formula  is only a  good approximation to the true
mass-spin relation for small  spins and small mass differences $M(\omega)-M$.
This observation might be relevant for baryon phenomenology in the Skyrme
model.
There  a non-relativistic  formula like (\ref{rigid}) is used to calculate the
theoretical predictions for
the masses of the  nucleons and the $\Delta$ particle. However, the nucleon
mass is about
$10 \%$ larger than the mass of a  Skyrmion and  the $\Delta$
is about $40 \%$ heavier than a Skyrmion. Our calculations indicate
 that  the formula (\ref{rigid}) is a poor approximation for a  relative  mass
difference  as large as $40 \%$ and
that  it will generally  give too large a  value for
the mass of a spinning soliton  at  a given angular momentum.

\subsection{The case $\omega > \mu$}

When $\omega > \mu$   the equation (\ref{spinBe}) is the (unmodified) Bessel
equation of first order, all solutions of which are oscillatory  for large $r$.
Thus, in terms of $k=\sqrt{\omega^2 - \mu^2}$
we can write the asymptotic form of the solution in terms of the  Bessel
functions of  first and second kind
\bea
J_1(kr) =-{1\over k} {d J_0 \over dr}(kr) \sim \sqrt{2\over \pi
kr}\sin(kr-{1\over 4}\pi)
\eea
and
\bea
Y_1(kr)=-{1\over k}{dY_0 \over dr}(kr) \sim -\sqrt{2\over \pi kr}
\cos(kr-{1\over 4}\pi).
\eea
Both these functions may occur, so we write the asymptotic form
$\bvphi^{\omega}$ as
\bea
\label{waves}
\bvphi^{\omega}(t,\bx) = -{k\over 4}(p Y_1(kr) + q J_1(kr))\left(
\begin{array}{l}
\cos (\theta-\omega t ) \\
\sin(\theta -\omega t) \\
0 \end{array} \right),
\eea
where $q$ and $p$ are constants whose meaning we will explain later.
A hedgehog field with the  asymptotic form
(\ref{waves}) has infinite  energy  (\ref{mass}) and is thus rather
unphysical.
Nevertheless  this solution has a natural interpretation in terms of
the dipole model which we will give  below.

First, however, we want to study the time evolution of a  baby
Skyrmion which is given an $\it initial$ angular velocity $\omega >\mu$.
To investigate this question we have solved the field equations (\ref{EL})
numerically with initial values
$\bphi (t=0) =\bphi^{(1)}$ and $\dot \bphi(t=0) = -\omega \bn \times
\bphi^{(1)}$.
for $\omega=0.5$ and $\omega =0.9$. The grid for our simulations is a square
of $250\times 250$ points, extending in both the $x$ and $y$ direction from
$-25$ to $25$.
 At the boundary we  set the field
to  the vacuum value $\bn$ and absorb any incident kinetic energy.  For both
of the initial values of $\omega$ we find that the baby Skyrmion
radiates. In figure 3 we display the field of the baby Skyrmion whose initial
angular velocity
was $\omega =0.5 $,  10 units of time after the start of the simulation. The
picture clearly shows the sort of spiral pattern which is familiar from  dipole
radiation in linear relativistic field theories
\cite{pictures}.

The radiation carries away both  energy and angular
momentum. As a result
the baby Skyrmion slows down until the angular velocity has dropped
to a value below $\mu$ in both simulations. Figure 4 shows how the total energy
 for the two
simulations  decreases with time. We have also checked that after 1000 units of
time
the field has settled down to a uniformly rotating field of the form
(\ref{spin}) and extract the angular  frequency. In the simulation where
initially
$\omega =0.5$ we now find $\omega \approx 0.28$ and  in the simulation where
initially
$\omega =0.9$ we now find $\omega  \approx 0.3$.

It  is instructive to interpret both the solution of the hedgehog form and the
numerically found solution with the spiral pattern
shown in figure 3 in terms of the dipole
model.
For  this purpose it is best to  combine the asymptotic fields
$\varphi^1$ and $\varphi^2$ into the complex field $\Phi = \varphi^1 +
i\varphi^2$.  Then, the Klein-Gordon equation (\ref{spindip}), with
dipole moments given by (\ref{dips}) (where now $\omega > \mu$),
can be written
\bea
\label{complex}
(\Box + \mu^2)\Phi(t,\bx) = -p e^{-i \omega t}(\partial_1 +
i\partial_2)\delta^{(2)}(\bx).
\eea
To solve this equation we separate the time dependence in the form
\bea
\Phi(t,\bx) = e^{-i\omega t}g(\bx),
\eea
so that $g$ has to satisfy the static equation
\bea
\label{staticc}
(\Delta + k^2)g(\bx) = p(\partial_1 +i \partial_2)\delta^{(2)}(\bx),
\eea
with $k$ as defined above.
Next we need  suitable Green functions $G$ of the Helmholtz equation
in two dimensions, normalised so  that
\bea
\label{Helmholtz}
(\Delta +k^2)G(kr) = \delta^{(2)}(\bx).
\eea
A solution which describes an `outgoing' wave at  infinity can
be expressed in terms of the  first Hankel function
\bea
G^+ (kr)= {1\over 4i} H_0^{(1)}(kr) \sim -i\sqrt{1\over 8\pi kr}
e^{i(kr-{\pi\over 4})},
\eea
and a solution which describes an `incoming' wave at infinity
is given in terms of the second Hankel function
\bea
G^-(kr)= {i\over 4} H_0^{(2)}(kr) \sim i\sqrt{1\over 8\pi kr}
e^{-i(kr-{\pi\over 4})}.
\eea
A  particular real solution  of (\ref{Helmholtz}) is thus given by
\bea
{1\over 2}(G^+(kr) + G^-(kr)) = {1\over 4} Y_0(kr),
\eea
but to obtain the general solution we  should add an arbitrary
multiple of the  real solution of the homogeneous Helmholtz equation
\bea
{1\over 2}(G^+(kr) - G^-(kr)) = {1\over 4} J_0(kr).
\eea
 Thus,  converting  to polar coordinates
\bea
\partial_1 +i\partial_2 = e^{i\theta}({\partial \over \partial r} + {i\over
r}{\partial \over \partial \theta})
\eea
and choosing  the  real Green function ${1\over 4}Y_0(kr)
+{q\over 4 p} J_0(kr)$ for some  real number $q$ we obtain
a solution  of (\ref{complex})
\bea
\Phi^{\mbox{\tiny r}}(t,\bx) = -{k\over 4}(p J_1(kr) +  q Y_1(kr))e^{i(\theta
-\omega t)},
\eea
which is just the asymptotic form  $\bvphi^{\omega}$ (\ref{waves})
 of the hedgehog field written in complex notation.
Thus the hedgehog spinning at $\omega>\mu$ represents a solution
 with a radiation field that consists of both incoming and outgoing
radiation. This is the physical origin of the infinite energy of the
hedgehog solution.

It is not difficult to guess which Green function will lead to
a solution of (\ref{complex}) displaying the spiral pattern
observed in our simulation of the spinning baby Skyrmion.
Consider the solution  constructed from the
Green function $G^{+}$.
Using
\bea
 {d H_0^{(1)}\over dr} (kr) = -k H^{(1)}_1(kr)
\eea
that solution is
\bea
\Phi^+(t,\bx) =i{k p \over 4}e^{i(\theta -\omega t)}H^{(1)}(kr)
\sim p \sqrt{k\over 8\pi r}e^{i(kr +\theta -\omega t-{\pi \over 4})}.
\eea
Remembering that the real and imaginary part of  $\Phi^+$ should be
interpreted as the first two components of the asymptotic field $\bvphi^+$ of a
spinning baby Skyrmion we see that, for sufficiently large $r$
and a fixed value of $t$, the direction of $\bvphi^+$ is constant along
the spirals $kr = -\theta$.
This is precisely
the spiral  pattern   we  observed in the numerical simulation of a    baby
Skyrmion  spinning  with $\omega>\mu$.
The dipole picture shows that it can  be accounted for in terms of the Green
function $G^+$.

The dipole picture  can  be used to make sense of  many of the  qualitative
properties of a spinning baby Skyrmion. In principle  one could
also check whether the energy loss through radiation  plotted in figure  4 can
be quantitatively   modelled in terms of dipole radiation. However,
 the centrifugal effects in spinning baby Skyrmions change the dipole
strength, which therefore depends on the baby Skyrmion's angular frequency.
This considerably complicates the calculations and we therefore  have not
pursued this path.

\section{The Dipole Model for the Interaction of   Baby Skyrmions}

In section 3 we saw that a baby Skyrmion acts like the source of a doublet
of scalar dipole fields. In \cite{PSZ} it was shown, assuming a certain
superposition procedure for well-separated baby Skyrmions, that  a baby
Skyrmion also reacts to the field
of a distant baby Skyrmion like a doublet of scalar dipoles. There is a similar
correspondence
between a  superposition procedure and a linear model for the forces
between solitons  the Skyrme model. In \cite{S} it was shown that the  product
ansatz   in the Skyrme model (without the pion mass term)  leads to
the same forces between well-separated moving and spinning Skyrmions as
 a simple dipole model for  Skyrmions, provided relativistic corrections
such as retardation effects  are
included in both approximations. In this section  we will ignore such
relativistic effects and describe the dipole model for
 slowly moving baby Skyrmions.

Thus consider two well-separated baby Skyrmions, the first centred at
$\bR_1$ and rotated relative to the standard hedgehog $\bphi^{(1)}$ by
an angle $\chi_1$ and the second centred at $\bR_2$
and  rotated by an angle $\chi_2$. From \cite{PSZ}  we know that, at large
separation, the leading term in the potential describing the interaction
of two baby Skyrmions is the interaction energy of
 two  doublets of scalar dipoles in the plane, one
situated  at $\bR_1$  and the other at $\bR_2$ such that $|\bR_1 -\bR_2|$
is large compared to $1/\mu$. The dipole moments of the first dipole are
$\bp_a$, $a=1,2$, where
\bea
\bp_1 = p(\cos\chi_1,\sin\chi_1) \qquad \bp_2= p(-\sin\chi_1, \cos \chi_1)
\eea
and the dipole moments of the second are $\bq_a$, $a=1,2$, where
\bea
\bq_1 = p(\cos\chi_2,\sin\chi_2) \qquad \bq_2= p(-\sin\chi_2, \cos \chi_2).
\eea
Then, if the dipoles $\bp_1$ and  $\bq_1$ and the dipoles $\bp_2$ and $\bq_2$
interact via a scalar field obeying the Klein-Gordon equation with mass $\mu$,
the interaction energy between the  doublets  $\bp_a$ and $\bq_a$ is
\bea
W= \sum_{a=1,2}{1\over 2 \pi}
(\bp_a\cd\nabla)(\bq_a\cd\nabla)K_0(\mu R),
\eea
where $\bR = \bR_1 -\bR_2$ and $R= |\bR|$.
Introducing also the relative angle $\psi = \chi_1 -\chi_2$ one finds
\bea
W (\psi,R)= {p^2\over \pi} \cos \psi \Delta K_0(\mu R) =
{p^2 \mu ^2 \over \pi} K_0(\mu R)\cos\psi.
\eea
In the last step we have used (\ref{Green}) and  have omitted the
$\delta^{(2)}$-function term   because we are only interested in large
separations $R>1/\mu$.

Thus we   obtain the first prediction of the dipole model. The force between
 two baby
Skyrmions depends on their relative orientation. In particular  two
baby Skyrmions in the same orientation repel each other;  if one is rotated
relative
to the other by $90^{\circ}$ there are no static forces; if one is rotated
relative to the other by  $180^{\circ}$ the forces are attractive.
The forces  always act along the line joining the two baby Skyrmions,
but in addition  there is a  torque which tends to rotate the relative
angle to $180^{\circ}$. In analogy with the terminology used in discussing
Skyrmion dynamics we call this configuration the most attractive channel.

To obtain a more quantitative picture we must take into account the
mass and the moment of inertia of the baby Skyrmions. We assume that
the rotations of the individual baby Skyrmions are sufficiently slow
so that we can approximate the functions  $M(\omega)$ and $\Lambda(\omega)$
by the constants
 $M_0$ and $\Lambda_0$.
Hence our  model for  the asymptotic dynamics of two baby Skyrmions
has the Lagrangian
\bea
L_{\mbox{\tiny dipole}}={1\over 2}M_0\dot{\bR_1}^2 +{1\over 2}M_0\dot{\bR_2}^2
+{1\over 2} \Lambda_0 \dot \chi_1^2  +{1\over 2} \Lambda_0 \dot \chi_2^2
- W(\psi, R).
\eea
In fact the centre of mass position $\bS = (\bR_1 + \bR_2)/2$ and the
angle $\chi =(\chi_1 +\chi_2)/2 $ are cyclical coordinates and decouple
from the remaining coordinates.
Thus we work in the centre of mass frame and  set $\chi =0$ and
$\bS =0$. Moreover  we can introduce polar coordinates $(R,\phi)$ for
the relative position vector $\bR$. Then $\phi$ is also a cyclical coordinate
and it is consistent to set $\dot \phi =\phi =0$. Then we obtain the
dynamical system  with equations of motion
\bea
\label{ode}
{1\over 2} \Lambda_0 \ddot{\psi} &=&{p^2 \mu^2 \over \pi}\sin \psi
K_0(\mu R) \nonumber \\
{1\over 2} M_0       \ddot{R} &=& {p^2 \mu^3 \over \pi}\cos \psi
K_1(\mu R).
\eea

These equations  can easily be solved numerically, and in the next section we
will compare the soliton trajectories predicted
by  them with those calculated from the full field equations
(\ref{EL}).  Some readers may then find  it useful to  think of the equations
(\ref{ode})
in terms of  the coupled motion of a pendulum and a point particle.
More precisely the angle $\psi$ may be thought of as characterising
the angular position of a physical pendulum. Then the first equation in
(\ref{ode}) specifies  the torque acting on
the pendulum: it vanishes at the stable equilibrium point $\psi=\pi$
and the unstable equilibrium point $\psi= 0$ and is maximal when $\psi=\pi/2$.
Moreover the strength of the torque depends on the `external' parameter
$R$ and decreases with increasing $R$.
The second equation in (\ref{ode}) can be interpreted as the  equation for
the  linear motion of a point particle with position  $R$. The force acting
on the particle depends on its position,  its strength decreasing with
increasing $R$, but is also controlled by the `external' parameter $\psi$.
When $\psi =\pi$ (the pendulum's stable equilibrium) the force is attractive,
tending to decrease $R$; when $\psi=0$ (the pendulum's unstable
equilibrium) the force is repulsive, tending to  increase \nolinebreak  $R$.

\section{Numerical Simulations}

All the simulations of the field equations
 to be discussed in this section  take place on  a   square grid of $250\times
250$ points, extending in both the $x$ and $y$ direction from  $-25$ to $25$.
The initial configurations  are constructed from two  baby Skyrmion fields
using the superposition procedure referred to above; thus,  these
configurations can be characterised by giving the
individual baby Skyrmions' positions and orientations.
 The  baby Skyrmions'   centre of mass position and the overall iso-orientation
 are  immaterial for the dynamics, and we  take the
former to be at the grid's origin and usually choose the latter so that the
individual baby Skyrmions'  orientations are equal and opposite.
For each simulation we will specify the  initial  relative position and
the initial relative  orientation, denoted $\psi_0$.  We will also consider
initial
conditions where both baby Skyrmions have some non-zero initial velocity.
We then work in the centre of mass frame, so that the baby Skyrmions'
velocities are equal and opposite, and we include the effect of Lorentz
contraction in our initial  configuration. Unless specified otherwise
we will  consider
initial velocities along the $x$-axis, so generically one  baby Skyrmion
is initially in
the  half plane $x>0$ with velocity  $(-v,0)$ and the  other in the
half plane $x<0$ with velocity $(v,0)$, where $0\leq v<1$.

\subsection{Scattering from  Rest}

Suppose the  two  baby Skyrmions are initially at rest and well-separated,
one centred  at $(10,0)$  and the other at $(-10,0)$.  We  have calculated
 the time evolution of the corresponding field configuration for
 a variety of initial orientations $\psi_0 \in[0,\pi]$ and in figure 5.a)
we plot $R/2$, the separation of either baby Skyrmion from the centre,
as a function of time. We only show the time evolution until the
baby Skyrmions collide - the actual collision will be discussed in
the next section.  The qualitative features of the time evolution
can easily be understood in terms of the dipole model.

Consider for example the motion  when $\psi_0 =\pi$.
Then the  baby Skyrmions are already in the most attractive channel
and   remain there; hence
there is  an attractive force between their centres throughout
their interaction. The plot of the actually observed time evolution
of $R/2$ shows exactly such an accelerating motion.
When $\psi_0=\pi/2$ the initial torque  is maximal, but the initial
 force vanishes. Thus the  relative orientation $\psi$ swings
through the attractive channel $\psi=\pi$, at which point the force
between the baby Skyrmions is maximally attractive, but then overshoots
and approaches $\psi=3\pi/2$. While $\psi$ is close to this value
the force between the baby Skyrmions is again very small or zero,
so that we expect the separation parameter $R$ to be a linear function
of time here. This is precisely what we see in figure 5.a).
When $\psi_0$ is decreased further the force between the baby Skyrmions
is initially repulsive.  The baby Skyrmions  move apart but at the same
time $\psi$ increases so that some time later the baby Skyrmions
are in the most attractive channel. The force is now attractive and
the baby Skyrmions approach each other again. As the relative orientation
oscillates the baby Skyrmions experience alternating attractive and
repulsive forces and thus perform the oscillatory motion most clearly seen
in the trajectory for $\psi_0=0.3\cd\pi$.
Finally the baby Skyrmions get trapped in the attractive channel  and  collide.

When $\psi_0$ is decreased further the initial repulsive force  increases and
as a result  the
baby Skyrmions'  separation
may initially increase so rapidly that the attractive force which the
baby Skyrmions experience  once they are in the attractive channel is
too weak to invert their relative velocity. The baby Skyrmions then
escape to infinity, which in our simulations means that they hit the
boundary of the grid. We have investigated  boundary effects by
sending a single baby Skyrmion towards the boundary with
velocity $v=0.1$ and find weak repulsive forces  when the baby
Skyrmion is approximately $8$ units away from the boundary. The relevant
boundary  for the present simulation is at $x=\pm 25$, so we interpret
simulations where $R/2$  becomes larger than $15$ as   `escape to
infinity'.  This happens for $\psi_0 \leq \pi/4$.
 In our simulations the smallest
value of $\psi_0$ for which the baby Skyrmions ultimately collide is
$\psi_0 =0.275\cd \pi$, for which the trajectory is also shown in figure 5.a).

We have also solved
the equations (\ref{ode}) numerically for a range of initial values $\psi_0$,
always setting $R(0)=20$,
$\dot R(0)=\dot \psi(0) =0$.
Comparing these solutions with the corresponding
 trajectories found in
our simulations of the full field equations we find  qualitative agreement
in all cases, but quantitative agreement only for the  first part of the
trajectories (typically $0\leq t  < 100$).
Moreover,
the dependence of the
solutions of (\ref{ode}) on $\psi_0$ is   exactly as found in the field theory.
 There is a critical value
$\psi_c$ such that for  $\psi_0 \in (\psi_c,0]$,
$R(t)$  tends to infinity as $t \rightarrow \infty$, and for $\psi_0
\in (\psi_c,\pi]$, $R(t)$ approaches zero (where the equations are singular)
after a number of oscillations which becomes arbitrarily large as $\psi_0
\downarrow \psi_c$.  Its numerical value is
$\psi_c \approx 0.288545$, which should be compared with the  value
$0.275$ found in the field theory.
In figure 5.b) we show  $R$ and $\psi$  as a function
of time for $\psi_0 =0.28855\cd \pi$. The qualitative features
of the interplay between the angular motion ($\psi$) and the linear motion
($R$) discussed earlier are clearly illustrated.

\subsection{Head-on Collisions}

The dipole model  only describes the  asymptotic part of the trajectories
discussed so far.
We have seen, however,  that for  $ \psi_0 >0.275\cd \pi$   the two baby
Skyrmions
ultimately  adjust their relative orientation so that they are in the
attractive channel, and collide head on.
In the next  set of simulations we study the outcome of such a head-on
collision in the attractive channel for a variety of different initial speeds.
We fix  $\psi_0 =\pi$ and place the baby Skyrmions at
$(7.5,0)$  and $(-7.5,0)$, giving them initial velocities $(-v,0)$ and
$(v,0)$ respectively, where $0.1 \leq v\leq 0.6$.

In all the simulations the baby Skyrmions merge into the ring-like structure of
the 2-soliton
solution and emerge at right angles to their initial direction  of motion.
After the scattering they move away from each other with their relative
orientation still in the attractive channel. This is the $90^{\circ}$
scattering
that  is now a familiar and apparently generic  feature of topological soliton
dynamics. However, in our model this scattering process is accompanied by
the emission of a large amount of radiation. In figure  6 we show the energy
distribution
immediately after the collision with  $v=0.6$: the  rings of radiation are
clearly
visible. The radiation carries away so much energy that the baby Skyrmions only
escape to infinity (for our purposes the boundary of the grid) for  $v\geq
0.46$. For smaller initial velocities the attractive forces
between the baby Skyrmions after the collision  pull them back and they perform
another head-on collision, again scattering through $90^{\circ}$ and   emerging
 along their initial direction of motion  and  in the attractive channel. The
second collision is again accompanied by the emission of radiation, so the
solitons  travel less far than after their first collision before they turn
round. This  process is repeated, but   now  the motion  remains close to the
ring-like 2-soliton solution
at all times.  The individual baby Skyrmions are no longer distinct, and the
motion  looks like an oscillatory excitation of the 2-soliton.
The emission of radiation, however, continues until the kinetic energy has
virtually disappeared.  The final
configuration   is numerically indistinguishable from the 2-soliton solution.

The scattering process described above is  much more radiative than any
observed
in previous simulations of lump scattering in  the ${\bf CP^1}$ model
\cite{sigma}
or in other  two-dimensional  versions Skyrme models with potentials different
from ours \cite{pot}. At first sight
this is surprising: the  radiation  in our model is  massive whereas
there are massless radiation modes in all the  comparable models mentioned
above.
However, our model is also the only one in which there are strong attractive
forces. Moreover these forces are short-ranged, so that the potential energy
functional $V$ (\ref{V}) has a large gradient  at configurations consisting of
two nearby
baby Skyrmions.  This means that the time evolution of a configuration in the
vicinity of  those points in the configuration
space is not adiabatic. It follows in particular that the adiabatic or moduli
space approximation proposed
for the Skyrme model in \cite{M2} is not suitable for describing  soliton
collisions in our model.

In the next set of simulations we keep the initial velocity $v$ fixed
 at $0.5$ but vary  the initial relative  orientation $\psi_0$
between $0$ and $\pi$.
Initially, the baby Skyrmions are again placed at $(7.5,0)$ and  $(-7.5,0)$.
To understand the ensuing interaction processes it is useful to note the
symmetries of the initial conditions. Recalling that in our conventions
the baby Skyrmions have equal and opposite initial orientations we first
observe
that  the initial configurations  are invariant under  the
reflection $P_x$ (\ref{Px}).
 Since $P_x$ is a symmetry of the
Lagrangian,  the configurations will remain invariant under that operation
during their
time evolution.
Hence, {\it assuming} that the baby Skyrmions separate after the
collision, we can deduce that they must separate along either the $x$-axis or
the $y$-axis. In other words, if scattering takes place, it must be scattering
through either $0^{\circ}$ (i.e. trivial), $90^{\circ}$ or $180^{\circ}$.
We also note that, if the baby Skyrmions separate along the  $y$-axis, the
requirement
of invariance under $P_x$ allows only two possibilities for
their individual orientations: either the standard orientation where, using
the conventions of figure 3, the fields point radially outwards, or the
standard
orientation rotated by $\pi$, where the fields point radially inwards.
Thus, after the scattering the baby Skyrmions either have the same orientation
and are  in the most repulsive channel or their orientations differ by $\pi$
in which case  they are in the most attractive channel.

When  $\psi_0=0$ or $\psi_0=\pi$ the initial configurations are  additionally
invariant under, respectively, the reflection
 $P_y$ (\ref{Py}) or the combination of $P_y$ with an iso-rotation by $\pi$. It
follows that the  configurations after the interaction must
have the same invariances. In particular for  $\psi_0=\pi$ we can predict
purely on the basis of symmetry that,    if the baby Skyrmions scatter
through $90^{\circ}$,  they must be equidistant from the origin after the
scattering and they must be in the most attractive channel.
Of course, this is precisely what we observed in our previous simulation.
For the other `special'  initial configuration, where  $\psi_0=0$,   we find,
however,  that the baby Skyrmions  scatter through $180^{\circ}$.
More precisely they  head towards each other and slow down until they come to
a halt at  a separation $R\approx 3$. Then they turn round and escape to
infinity along the
line of initial approach. The relative orientation
$\psi$ does not change at all  during this process.
How, within the constraints imposed by
 the symmetries described above, does the
scattering interpolate  between $90^{\circ}$ scattering and $180^{\circ}$
scattering as we vary $\psi_0$ from $\pi$ to $0$?

In our simulations we find  that there is an   interval $I^{90} =[\pi/6,\pi]$
such that the baby Skyrmions scatter through $90^{\circ}$ for $\psi_0 \in
I^{90}$
and a smaller interval $I^{180}=[0, \pi/20]$  such that they  scatter through
$180$ for $\psi_0\in I^{180}$.
However, as $\psi_0 \downarrow \pi/6$ the
baby Skyrmions emerge from the  $90^{\circ}$-scattering  with different speeds:
the baby Skyrmion moving in the positive $y$-direction moves faster
than the one moving in the negative $y$-direction. The
momentum  balance is restored by radiation which
is emitted  predominantly
along the negative  $ y$ -axis. This process is  presented schematically
in figure  7.b).

 When $\psi_0$ is in the remaining interval $I^{\mbox{\tiny capt}}
=(\pi/20,\pi/6)$  the two baby Skyrmions
do not separate after the collision but form the oscillatorily excited
state of the 2-soliton already encountered in the previous set of simulations.
However, this time the excited 2-soliton as a whole moves in the positive
$y$-direction, and radiation is emitted in the opposite direction
to restore the momentum balance. This process is sketched in figure 7.c).

The  notion of the most attractive channel is useful for summarising our
results. For $\psi_0 \in I^{90} \cup I^{\mbox{\tiny capt}}$ the baby
Skyrmions merge and   scatter through $90^{\circ}$;
the closer the two baby Skyrmions  are initially to being in the most
attractive channel the less radiation is emitted in the  scattering process.
For $\psi_0 \in I^{90}$ the baby Skyrmions therefore escape to infinity after
the collision; for $\psi_0 \in  I^{\mbox{\tiny capt}}$
they from a coincident configuration which we interpret as an asymmetrically
deformed  2-soliton solution. The increased radiation and its asymmetry
is due to that deformation. Finally, for $\psi_0\in I^{180}$ there is not
enough time  for the baby
Skyrmions' relative orientation to adjust before the collision, and as a result
the  force between
them is always repulsive. We have not discussed  negative values for $\psi_0$
separately here because  a  scattering process
with given negative $\psi_0$ is  related to the  processes with $-\psi_0$
by the  reflection  $P_y$.

Clearly the precise values of the boundaries   of the intervals $I^{90}$,
$I^{\mbox{\tiny capt}}$ and
$I^{180}$ depend on the  baby Skyrmions' initial separation and their
speeds.
We have  repeated the simulations just discussed with a different value of
the initial velocity $v$;
the  qualitative
features discussed  above and depicted in figure 7  are unaffected by such a
change.

\subsection{Scattering with non-zero Impact Parameter}

We have also investigated scattering processes with non-zero impact parameter.
 We  define
the impact parameter in the usual way  as the distance of closest of approach
between the
two soliton centres in  the absence of interaction, and denote it by
\nolinebreak $b$.

As in the previous simulations,
the nature of the   scattering  processes to be studied here  depends crucially
on the  value of $\psi_0$, but, again as before, there are basically two
regimes.
For $\psi_0$ smaller than some critical value (which depends on the impact
parameter and the initial velocity) the forces are repulsive
and the baby Skyrmions escape to infinity after scattering through some
non-trivial angle. When $\psi_0$ is larger than the critical value the
scattering
is more interesting: now the forces are attractive and baby Skyrmions may be
captured into a bound orbit. The orbiting motion is  accompanied by
emission of radiation. Since the qualitative nature of this process is
independent
of the precise value of $\psi_0$ we have investigated it in detail only for the
 attractive channel $\psi_0=\pi$.

For a quantitative  study  we have also concentrated on a fixed  initial speed,
 $v=0.4$,
and looked at  the dependence of the scattering on the impact parameter
$b$. We already know from previous simulations that the
 the baby Skyrmions will be captured in  a bound orbit  for $b=0$ (the
oscillatory excitation of the 2-soliton), but  it  is clear that the baby
Skyrmions will escape to infinity after the scattering
if $b$ is  made sufficiently large. Hence
there must be a critical impact parameter  which separates the two types
of scattering. From  our simulation we conclude that this
critical value is approximately $ 1.5$.

When the baby  Skyrmions  get captured in  a  bound orbit their centres trace
out ellipse-like figures whose perihelion rotates slowly
and whose diameter and  eccentricity decrease as energy is lost through
radiation.
This is best seen in   simulations  where the speed $v$ is small and the
impact parameter  is large. In figure 8  we show a part of the baby Skyrmions'
trajectories for the initial velocity  $v=0.1$ and the
impact parameter $b=  12 $.
It is not clear from our simulations
what the final state of such an orbiting motion is. It is quite possible
that there is a  non-radiating solution of the field equations where
two baby Skyrmions orbit each other with an angular frequency less than $\mu$.
Moreover, our analysis of the spinning baby Skyrmion can be extended to
2-solitons, showing that there  are finite-energy,  spinning 2-soliton
solutions of the hedgehog from
provided their angular frequency is less than $\mu$. However, contrary to
the  spinning baby Skyrmion these solutions may well be unstable.
In our simulations  of orbiting baby Skyrmions the kinetic energy
decreases  all the time but is never zero. It is  not clear whether  either of
the  periodic but non-radiating
solutions described above is eventually realised or whether the system
will continue to radiate until it reaches the static 2-soliton solution.

\subsection{The Effect of Relative Spin}

The last set of simulations we want to discuss addresses the effect of
relative spin on the
interaction of two baby Skyrmions. If one of two well-separated baby Skyrmions
is spinning and the other at rest the dipole forces between the two average
to zero over the spinning baby Skyrmion's period of rotation. Thus we expect
there to be no net force in this situation. This is indeed what we observe:
with the stationary baby Skyrmion placed at $(-7.5,0)$ and the other
one, spinning at $\omega =0.2$ (with the corresponding profile function)
placed at $(7.5,0)$ the relative separation oscillates around the initial
value $15$ with  a small amplitude  $\approx 0.1$  and angular frequency
$\omega$.

Next we investigate the combined  effect of relative spin and relative
motion of the baby Skyrmions' centres.
In order to see the influence  of {\it spin} rather than that  of the relative
{\it orientation} on the  interaction  we
should make sure  that the  interaction processes takes
much longer than one period of rotation.
Thus we keep the frequency of the spinning baby Skyrmion constant
at $\omega =0.2$ and send the two baby Skyrmions towards each other
along the $x$-axis  with a  small  speed $v$. Performing this simulation for
$v=0.02$ and
$v=0.05$ and a variety of values for $\psi_0$, we find that the baby Skyrmions
always repel each other and  scatter through $180^{\circ}$.
The distance of closest approach is  $>9$ and  the spinning Skyrmion's
angular frequency does not change during the  interaction.
This remarkable result can be understood in terms of the dipole model
as follows.
Since the angular velocity is large and the torque acting on the relative
orientation weak as long as the two baby Skyrmions are not too close
together we can  assume that angular velocity is essentially constant during
the interaction process. Thus, as the two baby Skyrmions approach each other
they experience  alternatively  an attractive and  a repulsive force
for approximately equal durations given by $1/(2\omega)$.
This leads to a sequence of alternating attractive and repulsive
impulses. Since the strength of the force between two baby Skyrmions
increases with decreasing $R$, it is clear that every attractive
impulse will invariably be followed by a {\it stronger} repulsive one. However,
 it may happen that at some point a repulsive impulse is strong enough to
invert the sign of the
relative velocity $\dot R$. Then that impulse
 will be followed by an attractive
impulse which is {\it weaker} than itself. Thus,
for a sufficiently small initial speed $v$  the baby Skyrmions will
always eventually repel each other  and escape to infinity.

This qualitative explanation  in terms of the dipole model can
be confirmed by numerical solutions of the equations (\ref{ode}).
 In figure 9 we show  the trajectories
  $R(t)$ and $\psi(t)$
for a particular set of  initial values.  The plots  show clearly that
$\dot \psi$ indeed remains essentially constant during the interaction
process, as assumed in our qualitative analysis of  the dipole model.
There is a further interesting effect, though. By carefully measuring $\dot R$
and $\dot \psi$ long after
 the interaction we find that, at $t>200$,  $\dot R >v$ and $\dot \psi <
\omega$ (the
difference here is very small). Thus rotational kinetic energy
has been converted into translational kinetic energy.

We observe similar effects in our simulations of the full field theory
which become more pronounced as $v$ is increased. We will describe
them for one particular simulation, where the baby Skyrmions  are again
initially at $(7.5,0)$ and $(-7.5,0)$, $\psi_0=\pi/2$ and $\omega =0.2$,
but now $v=0.055$.
 The scattering is again repulsive,  but now the  distance of closest approach
is only $\approx 7$.
At the point of closest approach the  spinning baby Skyrmion's angular
frequency decreases to roughly $0.1$ and the other
baby  Skyrmion begins  to spin  at that frequency, too.
The baby Skyrmions then move away from each other, each travelling  at the
considerably increased  speed $0.16$. One checks that the loss in rotational
kinetic
energy of ${1\over 4}(0.2)^2\Lambda_0$ is approximately   balanced by the gain
in translational kinetic energy of $(0.16)^2 M_0-(0.055)^2M_0$.

When $v$ is slightly  larger than $0.055$   the scattering
behaviour becomes sensitive to the precise value of the initial relative
orientation $\psi_0$. For some values of $\psi_0$
the baby Skyrmions repel as before, but for others
 their linear momentum
is large enough to overcome the repulsive barrier due to their
relative spin. Then the baby Skyrmions collide and form the oscillating and
radiating 2-soliton
configuration already encountered in previous simulations, but now
this configuration rotates as a whole. As in the simulation discussed at
the end of section 6.3 we are unable to decide  on the basis of our
numerical results whether all the kinetic energy will eventually be radiated
away or whether the system will settle down to a uniformly rotating 2-soliton
solution.

As $v$ is increased further the baby Skyrmions
 continue  to get captured after the collision, even for values of $v$ as
large as $0.5$. Finally, for $v=0.6$,
they scatter through $\approx 50^{\circ}$
and  escape to infinity.
The collision is again very radiative. Afterwards both baby Skyrmions  spin
 at the same angular velocity, whose numerical value is less than $0.1$.
 Thus the  radiation  carries away  spin as well as energy.
`Exotic' scattering  angles in head-on collisions like the one observed
here are a familiar feature of  lump scattering in the presence of overall
spin, see \cite{sigma} and \cite{Qlumps}.
Their occurrence is not surprising.
If one or both baby Skyrmions are spinning initially, the initial configuration
is, in general,  no longer  invariant under the reflection operations
discussed in section 6.2. Thus, in the presence of spin,  there is no reason to
expect
the scattering angle in a head-on collision to take on
special values like $90^{\circ}$ or $180^{\circ}$.

\section{Conclusion}

Thinking of baby Skyrmions  in terms of pairs of  orthogonal dipoles
has proven very useful for understanding
their dynamical properties.
The dipole  picture reproduces the qualitative features of  the asymptotic
field of a spinning
baby Skyrmion, both in the radiative regime where the angular frequency
$\omega$ is larger than $\mu$ and in the non-radiative regime $\omega<\mu$.
The dipole model is most successful when applied to  the  interaction between
well-separated baby Skyrmions. Here it gives an accurate, quantitative
description.
Moreover it
allows one to understand in simple qualitative terms
the repulsive nature of the
forces between  baby Skyrmions which are spinning rapidly  relative to each
other and whose centres move slowly relative to each other.
The dipole model also predicts  the possibility of   transfer of spin and
rotational kinetic energy from one baby Skyrmion to the other,  in agreement
with  our simulations of the full field theory. This `spin-exchange' scattering
is  rather reminiscent of the  electric
charge exchange in dyon scattering \cite{dyon}.

The dipole picture is based on the linearisation  of the non-linear  baby
Skyrme model. When  studying  the change in shape of a spinning baby Skyrmion
or the interaction of two baby  Skyrmions close together,   however,
non-linear effects dominate  and numerical methods
become indispensable.
Here   our  simulations also contain  a number of lessons  for the discussion
of Skyrmion
dynamics.

We saw that for a  rapidly  baby spinning Skyrmion  centrifugal  effects
lead to a linear  dependence of the mass on the angular momentum. This should
be contrasted with the quadratic dependence predicted by the non-relativistic,
rigid body treatment commonly used in the Skyrme model. Our results
underline the importance of a more careful treatment of centrifugal effects
when extracting the baryon spectrum from
the Skyrme model.

The  notion of the
attractive channel proved  useful  for understanding
a wide range of scattering processes. Even though the forces between baby
Skyrmions are  complicated due to their dependence on the relative  orientation
(as
it is  for Skyrmions)  the torque acting on the relative orientation is
so strong that interacting baby Skyrmions  tend to be in the attractive channel
by the time  they collide. Then they scatter through $90^{\circ}$ as if they
had been in the attractive channel initially. However,
 this process is the more radiative the further the baby Skyrmions were
initially from being  in the attractive channel.

Finally, our results illustrate the importance of radiation in
soliton dynamics. Our simulations show that
a particularly  large
amount of radiation  is emitted in a head-on collision of two baby Skyrmions.
This means that baby Skyrmions  only  escape to infinity after the interaction
when
the initial relative  speed or the impact parameter  are quite large.
Otherwise they get captured and, after further emission of radiation,
settle down to  the static 2-soliton configuration.
We have already pointed out that the radiative nature of this process is
probably due
to the large gradient of the potential energy functional at configurations
which describe two baby Skyrmions close together.
It  may even  be possible to relate
the large amount of radiation accompanying most soliton interactions in our
model to the fact that  the energies of the static soliton solutions are much
larger (about 50 $\%$) than the Bogomol'nyi bound (\ref{Bog}).
In any case, it is clear that the  adiabatic approximation  to
soliton dynamics proposed in  \cite{M2} would not be useful in  our model.
 Moreover, our results
show  that    the inclusion of the pion mass does not necessarily  make the
adiabatic approximation more
applicable, as is often claimed. In our model  soliton dynamics is much less
adiabatic  than
in comparable model with massless mesons.

\vspace{2cm}

\noindent {\bf Acknowledgements}

\noindent Parts of this paper were written while both BJS and WJZ were
visiting the Isaac Newton Institute in Cambridge. BJS acknowledges an SERC
research
assistantship.

\pagebreak

\pagebreak
\parindent 0pt
\centerline{\bf Figure Captions}
\vspace{1cm}
\centerline{\bf Figure 1}

Profile functions for a static  baby Skyrmion (bottom curve)
and  a spinning baby Skyrmion with angular frequency (from top to bottom)
0.316,0.3 and 0.2.

\vspace{1cm}

\centerline{\bf Figure 2}

a) The mass $M(\omega)$ of a spinning baby  Skyrmion  in units of  $4\pi$ as a
function of the
angular frequency

b) Mass-spin relationship  for a spinning baby Skyrmion. The crosses mark
pairs $(M,J)$ calculated via (\ref{mass}) and (\ref{J}) for the same value
of  $\omega$; both $M$  and $J$ are plotted in units of $4\pi$.
The solid line is a plot of  the function $\tilde M$ of $J$ (\ref{rigid});
here, too,
$\tilde M$ and $J$ are plotted in units of $4\pi$.

\vspace{1cm}

\centerline{\bf Figure 3}

 Plot of the field of a baby Skyrmion with initial angular frequency
$\omega=0.5$ at time t=10. At every lattice site in physical space we plot an
arrow of unit length
whose direction  is  that of $(\phi_1,\phi_2)$
(we identify the 1- and 2-axes in the target space $S^2$ with those in physical
space).
At the head  of the arrow we put a `$+$' if $\phi_3$ is positive and a
`$\times$' if $\phi_3$ is negative.
If $(\phi_1^2 + \phi_2^2) < 2\times 10^{-4}$ no arrow is plotted.
Thus the vacuum  is  represented simply by a `$+$'.

\vspace{1cm}

\centerline{\bf Figure 4}

 Total energy of  spinning baby Skyrmions with initial angular frequencies
$0.9$ (top) and  $0.5$ (bottom) in units of $4\pi$.

\vspace{1cm}
\vbox{
\centerline{\bf Figure 5}

a) Relative motion of two baby Skyrmions released from rest.
$R/2$ is plotted as a function of time for, from  top to bottom,
$\psi_0 =0.275\cd\pi,0.3\cd\pi,0.4\cd\pi,0.5\cd\pi$ and $\psi_0=\pi$.

b) Prediction of the dipole model for the relative motion of two baby Skyrmions
released from rest. $R/2$ and $\psi$ as  a function of time for
$\psi_0=0.28855\cd\pi$.

\vspace{1cm}}

\vbox{
\centerline{\bf Figure 6}

Total energy density  shortly after the head-on collision of two baby Skyrmions
with initial speeds $v=0.6$.

\vspace{1cm} }

\vbox{
\centerline{\bf Figure 7}

 Head-on Collisions

a) Sketch of baby Skyrmion velocities and radiation emitted  shortly after
scattering
with $\psi_0 =\pi$

b) Sketch   of baby Skyrmion  velocities and radiation emitted  shortly after
scattering
with $\psi_0 ={\pi\over 6}$

c) Sketch of  2-soliton velocity and radiation emitted shortly after
scattering with $\psi_0 \in I^{\mbox{\tiny capt}}$

d) Sketch of baby Skyrmion velocities  shortly after scattering with $\psi_0=0$

\vspace{1cm} }
\vbox{
\centerline{\bf Figure 8}

Trajectories of two baby Skyrmions   with initial speed
$v=0.1$  and impact parameter $b=12$

\vspace{1cm} }

\centerline{\bf Figure 9}
The effect of spin:
trajectories $R(t)/2$ and $\psi(t)/2 \pi$ calculated from the dipole model
with initial values $R(0) = 15$, $\dot R(0)=0.06, \psi_0=0.5\cd \pi$ and
 $\dot \psi(0)=\omega =0.2$

\end{document}